24.02.10

# Performance of wire-type Rn detectors operated with gas gain in ambient air in view of their possible application to early earthquake predictions

G. Charpak[1], P. Benaben[1], P. Breuil[1], E. Nappi[2], P. Martinengo[3], V. Peskov[3,4]
[1] Ecole Superieure des Mines in Saint-Etienne, France
[2] INFN Bari, Bari, Italy
[3] CERN, Geneva, Switzerland
[4] UNAM, Mexico

**Abstract**

We describe a detector of alpha particles based on wire-type counters (either single-wire or multi-wire) operating in ambient air at high gas gains ($10^2$-$10^3$). The main advantages of these detectors are: low cost, robustness and ability to operate in humid air.
The minimum detectable activity achieved with the multi-wire detector, for an integration time of 1 min, is 140 Bq/m$^3$, which is comparable to that featured by commercial devices. Owing to such features the detector is suited for massive application, for example for continuous monitoring of Rn or Po contaminations or, as discussed in this paper, in a network of Rn counters in areas affected by earthquakes in order to verify, on a large statistical basis, the envisaged correlation between the sudden Rn appearance and a forthcoming earthquake.

## I. Introduction

It is well known that Radon gas from environment, accumulating in buildings, especially in the basements, represents a significant risk to the human health. According to the US Environmental Protection agency [1] Radon is the second (after cigarette smoking) cause of lung cancer. The most dangerous, for the health, Rn decay products are $^{218}$Po and $^{214}$Po, emitting alpha particles of energies 6 MeV and 7.7 MeV, respectively. Therefore, a continuous monitoring of Rn concentration in public areas may have an important impact on health safety. Nowadays, there are several off-the-shelf detectors working on different principles and satisfying the main requirements for an effective monitoring of Radon in buildings. The most widely used detectors are based on the accumulative technique, i.e. on continuously recording doses during a time interval ranging from 2 to 90 days, for example employing charcoal.

In this paper, however, the main focus will not be on the application to health safety, which is quite well covered by the existing detectors, but on the possibility to investigate anomalous Radon concentrations as precursor of earthquake events. This study requires the use of a large number of fast sensors operating online.
In the last decade, some studies have shown the possibility to correlate elevated concentrations, in the soil or in groundwater, of gas Rn, or changes in the concentration, to the early prediction of earthquakes (see for example [2-9].
To confirm such studies with a larger statistics, a wide network of cheap, compact and highly sensitive Rn sensors is mandatory. To be statistically significant, several hundreds devices must be deployed in regions where an earthquake may potentially appear. Several commercial detectors can in principle satisfy the technical requirements, for example, ATMOS 12dpx [10] or RADIM3 [11], but unfortunately all of them are too expensive to be employed in massive applications. Their high cost is justified by their excellent performance, for example in term of spectroscopic response. However, for the purpose of measuring Rn concentration in several locations and determining the correlation between the different measurements such a good energy resolution is not required. For this purpose it is sufficient to record simultaneously signals above a given threshold, thus providing a reliable information about the Rn appearance and accumulation.

The key points for the implementation of this approach are low-cost, low power consuming sensors. The detector array must be equipped with a network of radio transmitters which broadcast signals to a headquarter, where the data are stored and analysed. In order to extend the battery life time, we suggest to perform the measurement for a few minutes each hour. This approach has already been successfully tested on several battery operated devices [12].

In [13], a cheap and simple gaseous device for the indoor detection of Rn was described. This detector consists of two RETGEMs (Resistive electrode gas electron multipliers) operating in ambient air in cascade mode with an overall gas gain of $10^3$-$10^4$. For the sake of completeness, we recall that the RETGEM is a hole-type structure manufactured from a G-10 sheet, 0.5-1mm thick, coated on both sides with a thin resistive (~0.5M$\Omega$/□) layers (which serve as electrodes) through which holes 0.4-1mm in diameter and 0,8-1mm pitch are drilled.. The use of hole-type detectors is fundamental since air, as well known, is not a good quencher. Numerous studies performed by various authors (for example [14]) show that "classical" gaseous detectors, like parallel-plate chambers and ordinary single-wire counters, do not work stably in air at high gas gains due to, among several reasons, strong photon feedback .

Opposite to "classical" gaseous detectors, hole-type sensors, including RETGEMs, feature cathodes geometrically shielded from the direct light emitted in the avalanche process thus enabling to run them at high gains even in presence of poorly quenched gases.

The role of the resistive electrodes is to make the RETGEM very robust, by protecting the detector and the front-end electronics against occasional sparks [15].

Moreover, in some applications, like the monitoring of Rn in the basement of buildings, it is necessary to operate the detector in humid air. Our tests show that a single-stage RETGEM can operate normally up to a humidity level of 30% whereas, a double-step RETGEM runs stably up to 70% relative humidity. At higher humidity, because of the charge leaking across the hole walls, the detector becomes noisy. Unfortunately, there are no simple solutions to this problem and hence a RETGEM detector can be used only for outdoor applications.

In this paper, we investigated the idea to use wire-type counters in humid air. The main interest to use a wire detector is that there is practically no limit on the size of dielectric interface between the cathode cylinder and the anode wire thus allowing to design a device capable to fully avoid surface leaks due to humidity. However, to use wire–type detectors, the photon feedback must be reduced and a reasonable long-term stability reached. The simplest solution would be to run the detector in ionization mode, at a gas gain of one, which guarantees stable operation. Unfortunately, as was shown in [13] the efficiency for the detection of alpha particles would be too low.

Since high efficiency is mandatory for monitoring earthquake, we investigated the possibility to increase the stability of the single-wire detector operated in proportional mode.

In the following, we present preliminary, but very encouraging, results indicating that a careful geometrical optimization allows both the single and multi-wire counters to operate in proportional mode, both indoor and outdoor.

**II. Feedback and detection efficiency**

Our early studies have revealed that the photon and ion feedback in gaseous detectors depends on several parameters, for example the feedback probability decreases with the strength of the electric field near the cathode [16,17]. This reduction can be rather significant and it is due to the increase of the so called "back diffusion" of secondary electrons created from the cathode in a weak electric field [17].
It was also observed that the probability of the photon feedback strongly depends on the avalanche emission spectra:

$$\gamma_{ph}(E) = \int Q(E, E_v) S(V, E_v) dE_v \qquad (1)$$

where $Q(E,E_v)$ is the quantum efficiency of the cathode as a function of the electric field E at the cathode and the photon energy $E_v$, $S(V,E_v)$ is the emission spectra of the Townsend avalanche and V is the voltage applied. Typically $Q(V,E_v)$ sharply (by orders of magnitude) increases with the photon energy therefore, in most cases, the UV part of the avalanche emission spectra gives the strongest contribution to $\gamma_{ph}$
Hence, the photon feedback can be strongly reduced if the emission spectrum of the avalanches is shifted towards longer wavelength.
  Another important discovery was that the number of emitted photons per unit of charge in the Townsend avalanche decreases with the electric field (see [18,19]). Therefore if the avalanche is developing in a high electric field the photon feedback will be lower compared to the case when the same gain is achieved in a weaker electric field.
Finally, it was shown [16] that the probability of the ion induced feedback $\gamma_+$ depends, not only on the electric field near the cathode, but also on the ionization potential $E_i$ of the gas and the work function $\varphi$ of the cathode material:

$$\gamma_+ = k(E)(E_i - 2\varphi) \qquad (2).$$

All these earlier studies suggest that, in a single-wire counter with a small anode diameter and a very large cathode diameter, the feedback process could be strongly reduced and eventually high gain obtained in air. Indeed, a thin wire creates a strong electric filed in its vicinity thus reducing the light emission per unit charge in the avalanche. Due to the large cathode diameter, the electric field near the cathode will be low and thus the back diffusion effect will be stronger. In addition, due to the large diameter of the cathode the UV photons from the avalanche with a wavelength < 185 nm will be absorbed in the air and this will further reduce the photon feedback, as the photon feedback depends on the photon energy, see formula 1. Finally, it is possible to select a cathode material with a high work function $\varphi$ to further suppress the feedback.

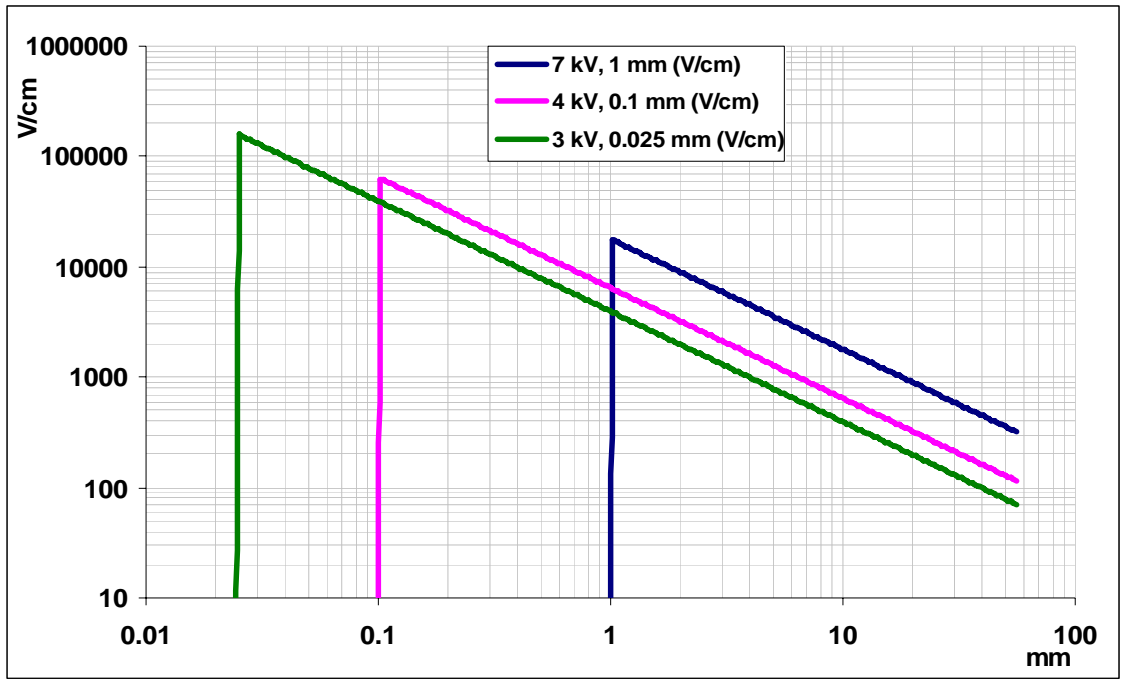

Fig.1. Electric field in a single wire counter as a function of the distance from the anode wire diameter: 25μm (green), 100 μm (rose) and 1000 μm (blue).

We can now try to answer to the crucial question: what efficiency to alpha particles can be achieved with a single-wire counter having a large cathode diameter and a thin anode wire?

Free electrons drifting in air experience electron attachment. At a radius r from the axis of the counter and having defined $r_c$ as the boundary within which the electric field is strong enough to cause gas amplification, the number of survived electrons will be:

$$n=n_0\{exp-\int_{r_c}^{R}(\eta dr)\} \quad (3)$$

where R is the radius of the cathode, $n_0$ is the number of primary electrons created by the ionizing radiation and $\eta$ is the attachment coefficient. As it was shown in [20-22], the dependence of the attachment coefficient on the electric filed E and the gas pressure p can be well described by the following formula:

$$\eta/p=A(E/p)+B(E/p)p+C(E/p)p^2 \quad (4)$$

where A, B and C are coefficients (function of E/p) which correspond to dissociative, three–body and four-body attachment to oxygen molecules respectively.

Calculations show [23] that the ratio $\eta/p$ first sharply drops as E/p increases, up to approximately 8 $Vcm^{-1}Torr^{-1}$, then in the region 8<E/p<28 $Vcm^{-1} Torr^{-1}$ it reaches a plateau and finally for E/p> 28$Vcm^{-1}Torr^{-1}$ it sharply drops again.

From the knowledge of $\eta/p$ as a function of E/p it is possible to calculate the probability, $n/n_0$, for a primary electron to escape to attachment. Such a calculation was performed together with measurements and results were reported in [23]. For a single-wire counter geometry, $n/n_0$ was found to be about 0.8 if the primary electrons drift a distance $\ell$ =1 mm towards the anode wire and $n/n_0 \sim 0.2$ for $\ell$ = 3mm. Thus, if the length of the ionization track L is larger than R or if the drift distance $\ell$ is smaller than $\lambda$, $\lambda$ being the mean free path of the electrons before they are captured, the avalanche near the anode wire will be triggered by free electrons.

The mean track length of alpha particles in air is L ~ 4cm therefore, if R ~ 1 cm, the probability to trigger an avalanche by primary electrons is close to 100%.

However, if R > L and $\ell$ the situation changes: most of the electrons will be captured by electronegative molecules during their drift and these negative ions will approach the anode wire where, in the region of the strong electric field, they will be stripped out so that finally the liberated electrons will initiate the Townsend avalanche. According to [24] there are three main mechanisms of electron detachment from the negative ions:
1) detachment due to the collision of negative ions with molecules followed by formation a new molecule.
2) detachment due to the collision of negative ions with metastable atoms and molecules.
3) detachment due to the collision of negative ions with molecules due to the high kinetic energy of ions acquired in the avalanche region.

Estimates show that, in an electric field of ~ $10^5$ V/cm, the negative ions can gain a kinetic energy close to 100eV, sufficient to knock out weakly bounded electrons from electronegative molecules, so this mechanism plays the major role in the initial stage of the avalanche formation.

During the development of the avalanche other mechanisms will contribute as well, the most important being the collision of negative ions with avalanche electrons which have a high cross section to free the attached electrons. The efficiency of the electron detachment via collisions with energetic ions depends not only on the strength of the electric field, but also on the size of the region $\Delta l$ in which the electric filed is high. Our previous experience with cascaded RETGEM [13] shows that high efficiency is achieved even when the electrons from the alpha tracks drift a distance much larger that the track length (up to 10 cm). In this detector, after the drift, negative ions enter the detector holes, where due to the high electric field, they experience electron detachment. Calculations with the Maxwell 3D electric field simulator show that in a 0.4 mm thick RETGEM ($\Delta l \approx 0.4$mm) the field can reach value of 40 kV/cm, for an applied voltage of 3kV [25]. In a single- wire detector with the anode wire diameter below $d_a<100$ μm the field near its surface is even stronger, close to100 kV/cm, see fig.1. However the region of high electric field is much narrower, $\Delta l \approx 50$ μm, see fig. 1. If $d_a>100$ μm the electric filed near the anode wire is below 40kV/cm , however $\Delta l$ becomes larger. Probably there is an optimal diameter of the anode wire at which the detachment has a maximum. Therefore it seems possible to efficiently detect alpha particles with a single-wire counter having a large cathode diameter and a small, optimized, anode diameter operated in avalanche mode without, or with strongly suppressed, feedback. Experimental results presented below fully confirm these expectations, allowing to build simple and cheap Rn detector working in ambient air.

## III. Detector lay-out and experimental set up

As pointed out earlier, RETGEM does not operate properly in humid air. In the single-wire design, this problem can be overcome implementing a dielectric surface between the anode wire and the cathode cylinder.

Two types of detectors have been developed and tested. The first is a simple single -wire counter extensively used for basic studies, called "basic detector", the second one is a more sophisticated detector, a multi-wire proportional counter (MWPC). Both are operated in avalanche mode.

The "basic" detector lay-out is shown schematically in fig.2. It is a single-wire counter with a stainless steel (SS) cathode cylinder of 60 mm diameter, anode diameter 100 μm.

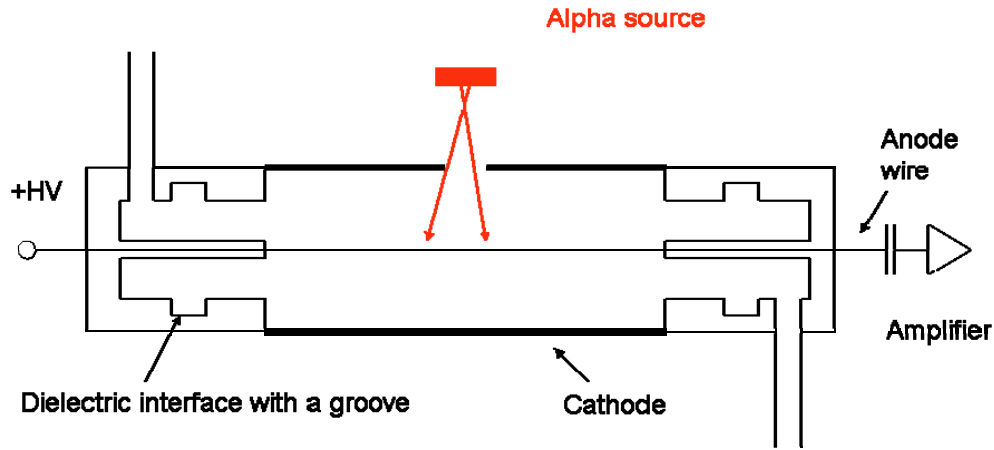

Fig.2. Schematic drawing of the single-wire counter working in humid air at high gain.

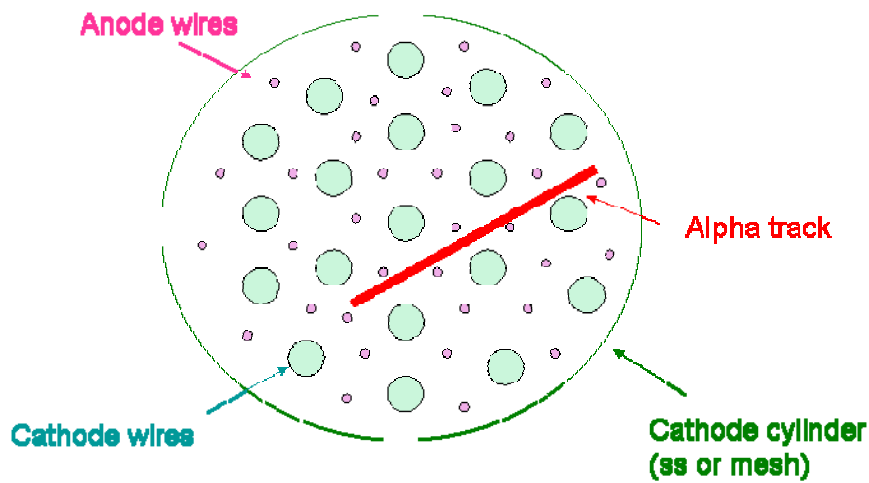

Fig.3. Schematic cross section of the MWPC sensor.

As explained above, this simple geometry features a reduced photon and ion feedback, enabling to operate the detector at relatively high gain in air. Other technical solutions have been implemented to prevent surface leaks in humid air. The most efficient solution is to create deep, rectangular and sharp grooves in the dielectric interface between the anode and the cathode electrodes. However, in order to prevent the current flow along the surface, the walls of the grooves must be perpendicular to the field lines.

In addition to the previous "basic" lay-out, for comparative studies, we have used counters with R=60mm and anode diameter of 25 μm and R=30mm, $d_a$=25 μm and 50μm.

In the measurements, a positive HV was applied to the anode whereas the outer cylinder was grounded. The signal from the anode wire was detected with a custom made charge-sensitive amplifier, able to deal with long–rise time pulses (due to the long drift time of negative ions) and, if necessary, it was additionally amplified by a research amplifier.

Subsequently, it was displayed on a oscilloscope and pulses counted by a scaler. In some high precision measurements, like the study of the sensitivity limit for Rn detection, the signal was analyzed and displayed with a Labview application whit time measurement capability and , if necessary, the suppression of undesirable pick up oscillations or pulses having untypical shapes like short pulses.

A schematic drawing of the multi-wire detector is presented in fig. 3.

It consists of a cylinder of 40 mm diameter, 50 cm long made of SS with several holes along the axis for the passage of alpha particles or a SS mesh with anode and cathode wires stretched parallel to the cylinder axis in an hexagonal pattern. This design actually is a copy of the detector described in [26] and was manufactured using the same technology. However each anode wire has protective grooves to minimize the charge leak across the surface, on both sides of the top and the bottom dielectric flanges. These flanges have several holes of 1mm in diameter to allow for the entrance of a collimated beam of alpha particles parallel to the anode wires. The cathode wires are made of SS of 1.5 mm diameter, also providing a rigid support mechanical structure during the assembly.

The anodes are gold coated tungsten wires 25μm diameter. The cathode wire pitch is about 8 mm and the anode wire pitch is about 5 mm. During the basic studies of this device, we have detected and analyzed signals from individual wires. In high precision measurements, in order to reduce the noise, the electronic circuit was the same as often used in experiments with MWPC: each anode wire was connected to its own amplifier channel which after the amplitude discrimination produces a standard square pulse, 1μs long. These pulses were sent in parallel to a simple majority unit generating an output pulse in presence of the coincidence of two or more inputs. The output pulses were counted using a standard scaler. During the measurements with alpha particles, only events with two or more wires above threshold within a few μs gate have been recorded and analysed.

Most of the measurements were performed in ambient air. However, some measurements were done flushing the detector with Ar. In preliminary studies a $^{241}$Am alpha source was used, positioned at different distances from the opening in the cathode cylinder. During the measurements with $^{220}$Rn or $^{222}$Rn air with traces of these radioactive elements was introduced in the detector. Their concentration was evaluated from the counting rate.

To crosscheck the measurements samples of air were also independently measured by the French company ALGADE [27].

**IV. Results**
**IV-1. Single –wire detector**

**IV-1a. Results obtained with alpha particles from $^{241}$Am**

Figure 4 shows the signals (displayed with a LabView program) produced by a $^{241}$Am source when the detector was filled with ambient air. The large pulses were produced by alpha particles whereas the smaller ones were caused by 60 keV photons. In fact, after covering the Am emission surface with a thin film of scotch, fully transparent to 60keV photons, only the small pulses were observed. They disappeared when the source was removed or shielded with a lead plate.

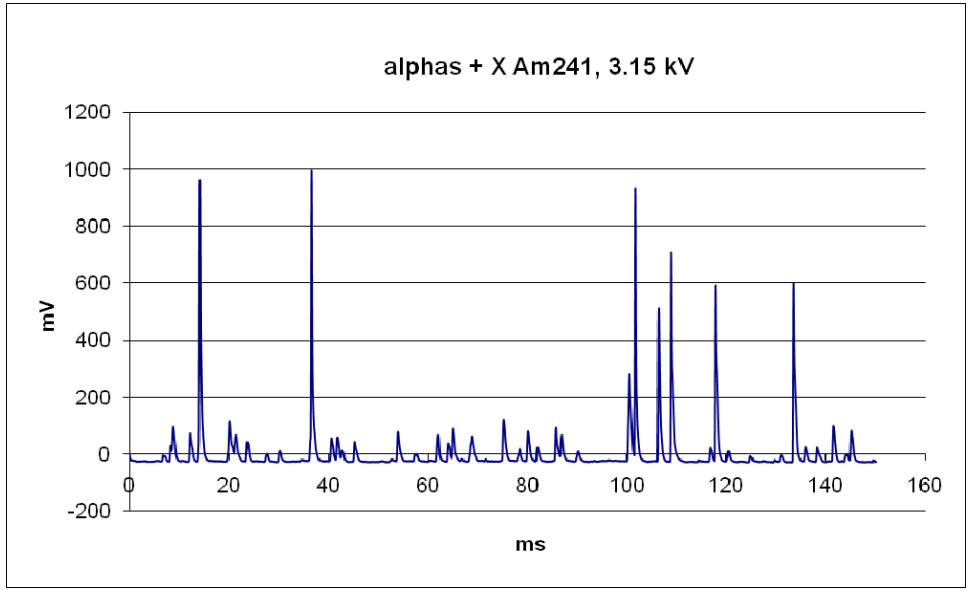

Fig. 4. Typical signals observed with the single–wire counter (basic lay-out) working in air and irradiated with a $^{241}$Am source

In figure 5 is plotted the mean values of the alpha signals as a function of the voltage applied to the single-wire detector, for a cathode diameter D=60mm and varying the diameters of the anode wire: 25, 100 and 1000μm.

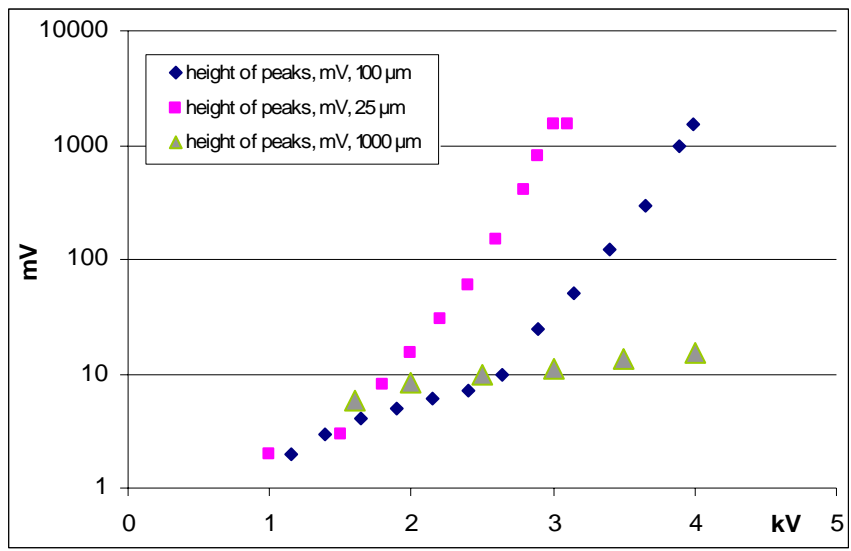

Fig 5. Mean signal amplitude produced by alpha particles vs. high voltage in a detector with cathode diameter of 60 mm and varying the anode diameter

For a small anode diameter, $d_a$=25 and 100 μm, at voltages $V_d$> 2kV the signal increases exponentially with the voltage whereas for 1mm diameter it varies very little in the range 2-4 kV.

As shown in [13], the later pulses are produced by positive and negative ions drifting toward the detector electrodes. For $d_a$=1mm there is no gas multiplication and the detector is operating in ionization modes. These small pulses were detected due to the high sensitivity of our custom made amplifier, capable to detect slow signals. However, as shown in [13], this mode of operation is very difficult to exploit in practice: the detector was extremely sensitive to vibrations and even under ideal conditions, when all vibrations were suppressed, the efficiency was not higher than 20%.

A smaller anode diameter enables gas multiplication, as a result, the signal amplitude can be easily increased by a factor 100, allowing the detector to be operated under realistic condition and with high efficiency.

In first approximation, the gain of a thin anode wire is given by:

$$G=\Delta A(V_d)_{thin}/ \Delta A(V_d)_{thick} \quad (5),$$

where $\Delta A(V_d)_{thin}$ and $\Delta A(V_d)_{thick}$ are the mean amplitudes measured with thin ($d_a$=25 and 100 μm) and thick ($d_a$=1mm) wires, respectively.

Similar dependences were observed with a 30mm cathode diameter counter. As an example, in fig. 6 are shown the mean amplitudes as a function of the voltage for sensor with D=30mm and $d_a$=100μm. The blue curve represents data obtained with alpha particles, whereas the pink curve shows data obtained with 60 keV X-rays. At $V_d$>3250V the pulse amplitude sharply increases with the voltage, indicating the appearance of the photon feedback. With our amplifier the photon feedback pulses following the primary one were integrated in one single pulse with higher amplitude.

From a practical point of view the most important characteristic of the Rn detector is not the maximum achievable gain, but its efficiency to alpha particles.

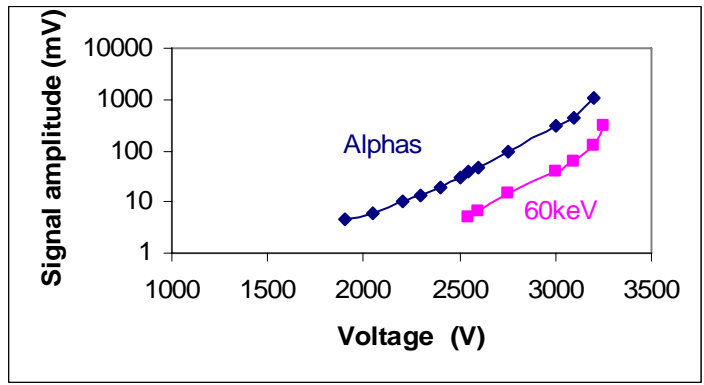

Fig. 6. The pulse amplitudes vs. the voltage measures with a single-wire counter having D=30mm and $d_a$=100μm. One can see that at $V_d$> 3250V the pulse amplitudes start deviating from the strait line in the log scale indicating the appearance of feedback processes. This effect was not observed in the case of detector having D=60mm and da=100 or 25 μm

To evaluate the efficiency, we performed comparative measurements with the same single-wire counter flushed with Ar. In Ar it is much easier to observe alpha signals at G=1, due to their shorter duration, and we could also operate the detector at elevated gas gain.

The efficiency was defined as the ratio between the counting rates measured in air ($N_{air}$) and in Ar ($N_{Ar}$):

$$\eta = N_{air}/N_{Ar} \quad (6).$$

Figure 7 shows the counting rates measured in air and Ar. The same counting rate was measured in the plateau regions of the curves, indicating that 100% efficiency to alpha particle was achieved in air. It is very important that the detector had a large plateau region when operated in air since it allows to perform precise measurements in a wide range of gain. Note that a nice plateau was obtained even in the case of 60keV photons, see fig. 8.

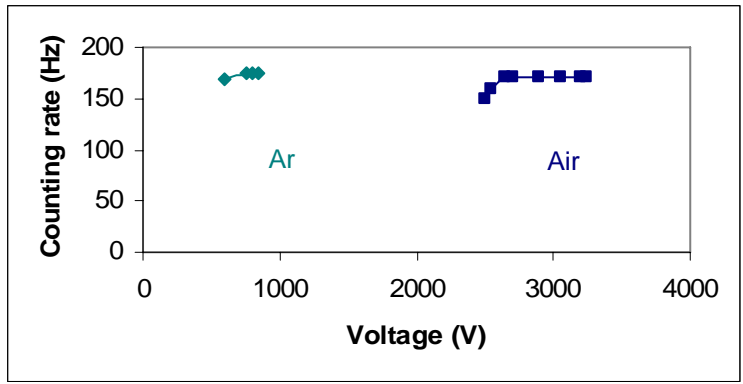

Fig.7. Counting rates vs. the applied voltage measured in Ar and in air under the same conditions (alpha particles, D=60mm, da=100 μm)

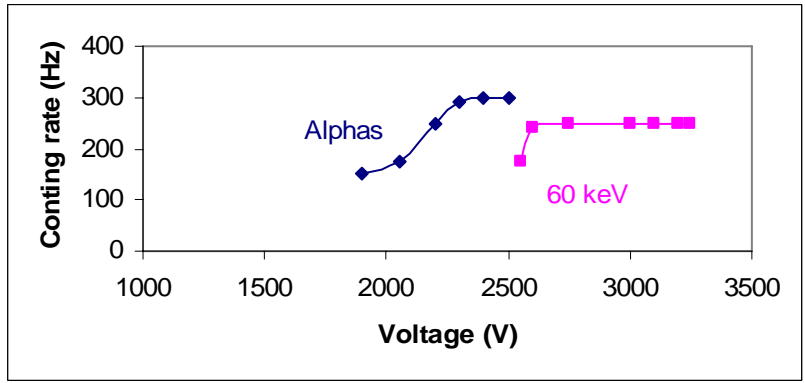

Fig.8. Counting rate vs. $V_d$ for alpha particle (blue) and for 60 keV photons (pink) measured with a basic design of the single-wire counter

A further very important observation is that the rate is almost independent of the humidity level: with an increase of the humidity from 0 to 80% the rate curves are only slightly shifted parallel towards lower voltages (on a few tenths of V). The same effect was also observed in [23]. In the humidity interval 80-100% the plateau region starts at 2.5kV and becomes slightly narrower, ending at 3.15 kV. The pulses detected with 100% humidity level are shown in fig.9. These measurements allowed to set the optimum operation voltage, about 3kV, for the measurements performed with Rn.

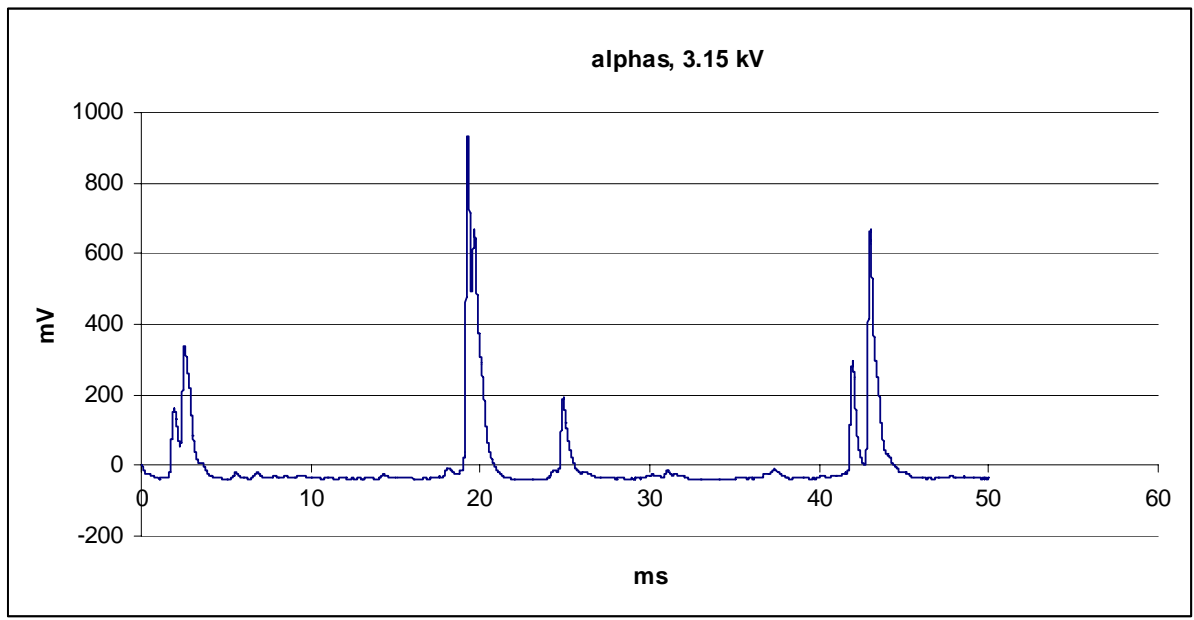

Fig.9 . Pulses produced by a $^{241}$Am in 100% humid air

## IV-1b. Results with $^{220}$Rn (Thoron)

The measurement of small Rn concentrations and, especially, the detection of fast changes in the concentration requires high efficiency coupled to a very low noise.
For a single-wire counter operated at high gain, the main source of noise are: cosmic rays, spurious pulses from the cathode and, in presence of high humidity, occasional avalanches along the surface, see [28]. Noise pulses can be discriminated from the alpha pulses by comparing the amplitudes and their width. Indeed, alpha particles produce almost $10^5$ primary electrons and ions in air. As it was already described above, most of the primary electrons will be rapidly captured by electronegative molecules, which will rather slow the drift towards the anode wire, where in the region of the strong electric field detachment process will occur and the liberated electrons will trigger the Townsend avalanche. This phenomenon produces a long pulse on the anode wire determined mainly by the drift time of the negative ions. In contrast, spurious pulses caused by the electron emission from the cathode, due to the ion recombination or electron jets [28], will produce short pulses, smaller in amplitude than alpha pulses. Cosmic radiation will produce series of small pulses induced by the arrival of clusters of negative ions generated by the clusters of primary electrons.
To verify these expectations, we have studied the shape of pulses produced by alpha particles, by electron emission and cosmic rays. In order to be as close as possible to the real conditions these studies were performed not only with the $^{241}$Am source, but also with $^{220}$Rn (Thoron).
The convenience to use Thoron is because it has about 1 min decay time, which on the contrary of $^{222}$Rn, does not provide long-term radioactive contaminations to the detector thus allowing careful studies of the pulses produce by the randomly distributed alpha particles tracks ($^{220}$Rn>$^{216}$Po+α) and by noise. Thoron was obtained in the following way: a sealed metallic box was filled with towel paper preliminarily impregnated with a solution of thorium nitrate and dried. Thoron was generated via so called "Thorium series" of radioactive decays:

$$^{232}\text{Th} > ^{228}\text{Rn} > ^{228}\text{Ac} > ^{228}\text{Th} > ^{224}\text{Ra} > ^{220}\text{Rn}\ldots \quad (7)$$

For the measurement purposes, a small portion of air was extracted from the box and then injected into the single-wire counter. As an example, fig. 10 shows the counting rate measured by the single-wire counter in presence of a small trace of Thoron.
The rate drops with time as expected from the Thoron decay time. A very small deviation from the theoretical curve is caused by the diffusion of a fraction of the injected Thoron.

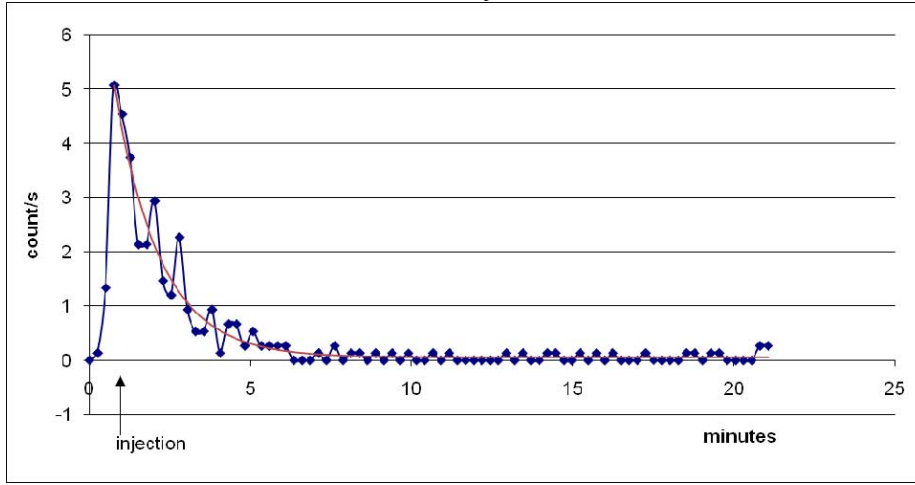

Fig. 10. Counting rate vs. time as measured by the single-wire counter in which air contaminated with Thorium was injected.

During the fist few minutes after the Thoron injection, the majority of pulses were produced by alpha particles. Their typical shapes are presented in figs 11a and b. They had either a smooth shape as shown in fig 11a or contained 1-2 additional peaks (fig 11b). The duration of the pulse was above 0.5ms. Ten-fifteen minutes after the injection the remained counting rate was mainly due to natural radioactive background and spurious pulses. All these pulses were shorter than 0.5ms and they had also a smaller mean amplitude (see fig. 12). This can be clearly seen from figs 13 and 14, comparing the pulses width and their amplitudes measured in pure air and in air contaminated with Thoron. Rejecting pulses shorter than 0.5ms and having amplitudes smaller than 70mV reduces the noise and hence increases the detector sensitivity to the presence of Rn. In our high precision measurements of Rn concentration such rejection was performed by the Lab view program.

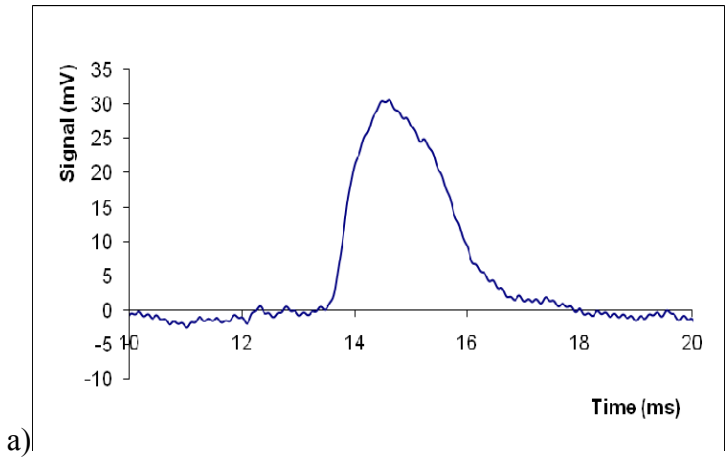

a)

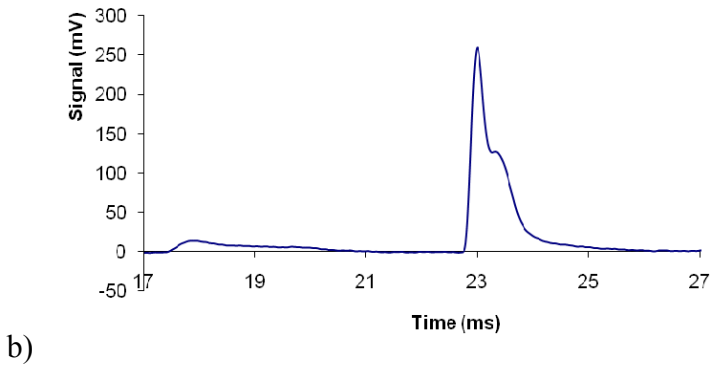

b)

Fig.11.Typical shape of pulses produced in the single-wire counter by Thorium: a) smooth pulses, b) pulses containing 1-2 peaks

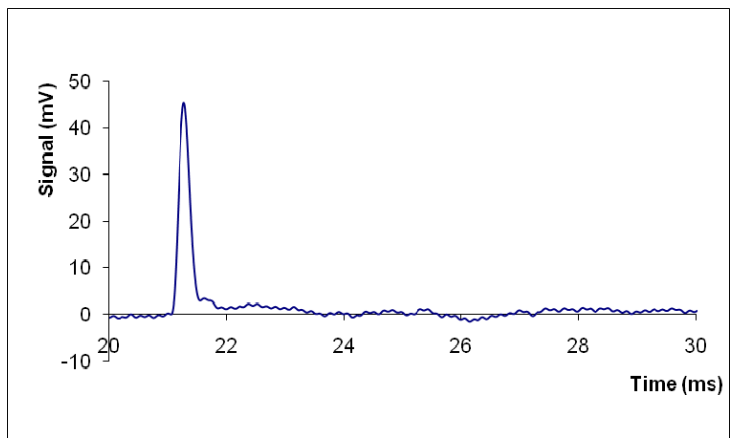

Fig. 12. Typical shape of noise pulses

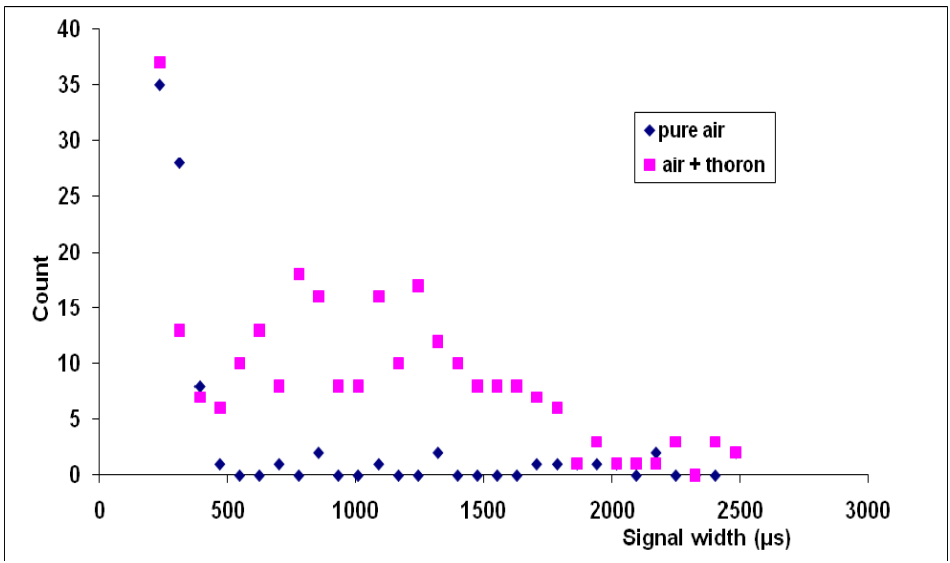

Fig.13. Distribution of the width of noise and Thoron induced pulses

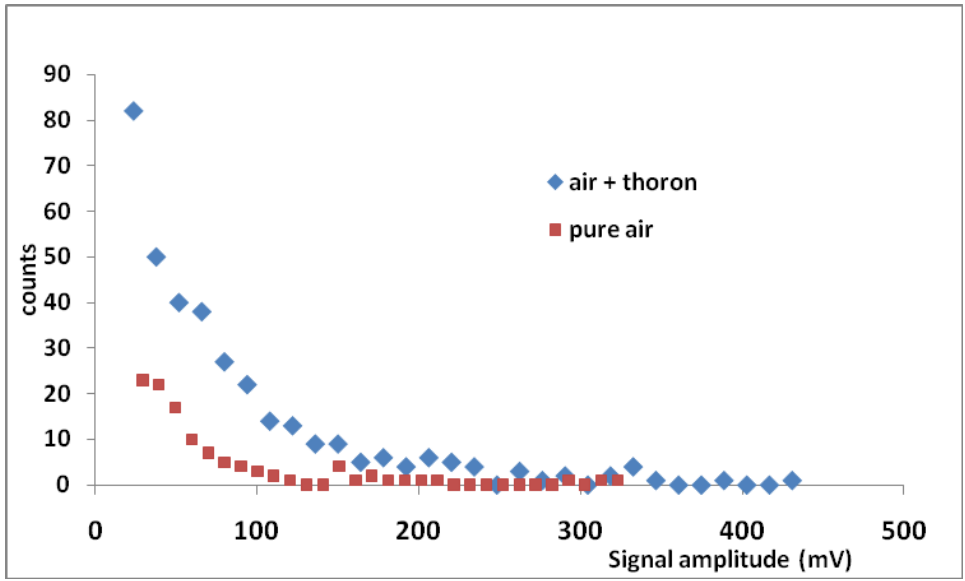

Fig. 14. Pulse height spectrum of Thoron and noise induced pulses

## IV-1c. Results with $^{222}$Rn

The final evaluation of the detectors was performed with $^{222}$Rn. Air with some trace of $^{222}$Rn was produced in the sealed metallic box containing about 50g of sandy loom with 1550 ppm of Uranium, obtained from one of the St. Etienne mines. The Rn was generated via the so called "uranium series":

$$^{238}U > {}^{234}Th > {}^{234}Pa > {}^{234}U > {}^{230}Th > {}^{226}Ra > {}^{222}Rn \qquad (8).$$

A volume of air of about 1l was extracted from this box and injected into the single-wire counter. Most of the measurements were performed with the "basic lay-out", however some comparative studies were also done with the single-wire counter having D=30mm and $d_a$=1mm and operated in pulse ionization mode.

Figure 15 shows the counting rate vs. time after the Rn was injected into the sensor. The counting rate N(t) first increases, mainly due to the Rn diffusion inside the chamber, and then the rate starts decreasing. Similar results, with a lower counting rate, were obtained with the ionization chambers, see figs 16 and 17. Since there are several decay channels for Rn producing alpha particles (see Fig.16), the total counting rate N(t) obeys the formula:

$$N(t) = N_m\{\exp(-t/T_0) + k_1 \exp(-t/T_1) + k_2 \exp(-t/T_2)\} + \ldots\} + N_0 \qquad (9),$$

where $N_m$ is the starting rate, proportional to the Rn concentration, $T_0, T_1$ and $T_2$ are the decay times of $^{222}$Rn, $^{218}$Po and $^{214}$Po respectively, $k_1$ and $k_2$ are coefficients reflecting the relative concentration of $^{218}$Po and $^{214}$Po and $N_0$ is the noise rate. Therefore the decay time of Rn itself can be clearly observed only 20-30 min after its injection inside the detector.

The main limitation to the sensitivity comes from the noise level. We observed that the counting rate of the noise pulses can be considerably decreased if the Rn contaminated gas remains in the detector for a relatively short time, less than 1 hour. If it remains for a longer period of time, the rate of noise remains high for almost two weeks, considerably longer than the Rn decay time.. This is due to the slow liberation of Rn trapped in the detector surfaces and inside the detector materials. Thus in high accurate measurements, to reduce $N_0$, we usually injected the Rn gas only for a short period of time, as shown in fig. 15, and then, when the measurements were finished the detector was flushed with clean air.

The minimum detectable activity (MDA) of the counter was determined from the known Rn concentration measured by the French company ALGADE [27], specialized in radioactive monitoring. The results obtained during a short accumulation time are summarized in Table 1. They revealed that the Rn detection efficiency $\eta_R$ of the basic design was 100% whereas in the case of the ionization chamber operating in pulse detection mode $\eta_R \sim$10%. The MDA of the basic design was 420Bq/m3, which is less than a factor 3 lower than the best commercial detectors, typically 150Bq/m3, see for example ATMOS [29].

If several detectors will be used in the same monitoring area, as suggested in the introduction, the real/ practical MDA will be improved by a factor $\sqrt{n}$, n being the number of sensors.

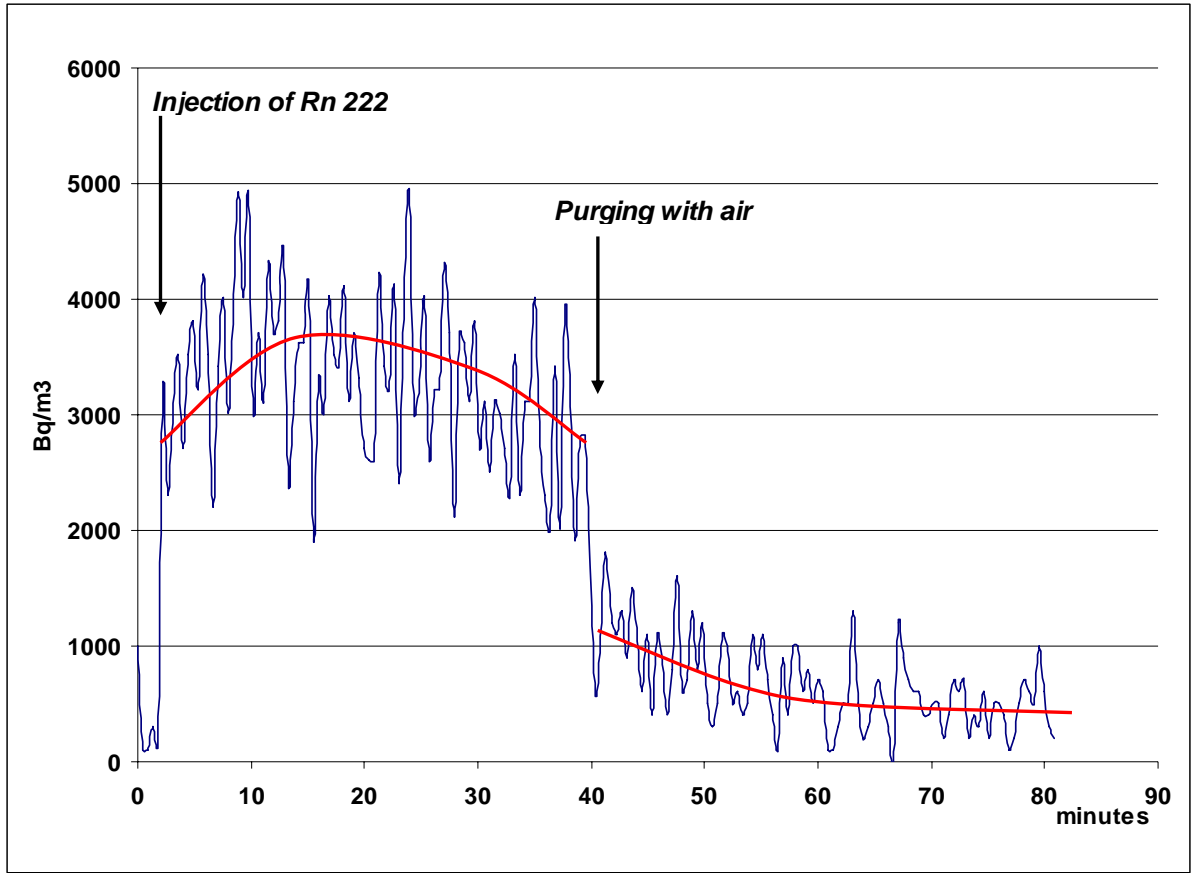

Fig.15. Counting rate vs. time after injection into the basic design (at t=2min) of air contaminated with $^{222}$Rn. At t=40 min the detector for a few second was flushed with clean air

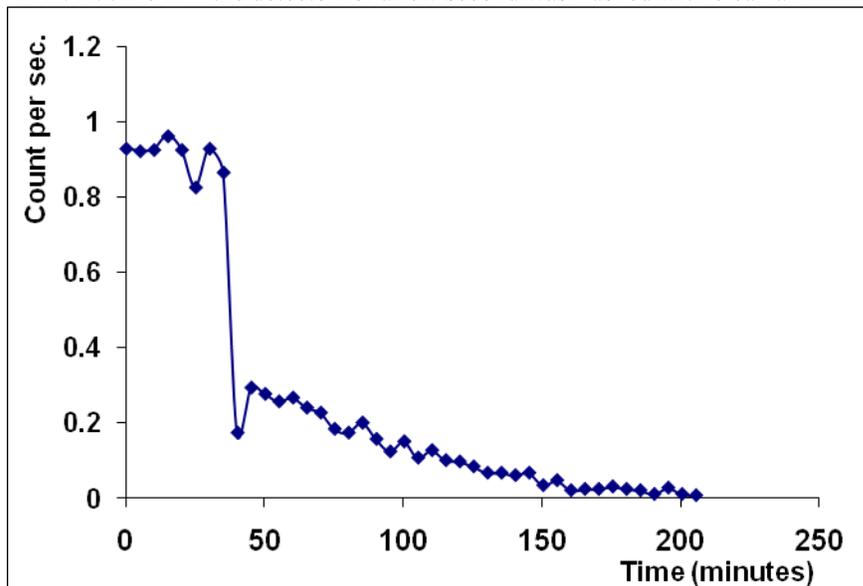

Fig. 16. Counting rate vs. time when the radon contaminated air was introduces (at t=0 sec) into the ionization chamber and at t=40 sec it was flushed with clean air. The fast decrease of the counting rate is mainly due to the Po decay

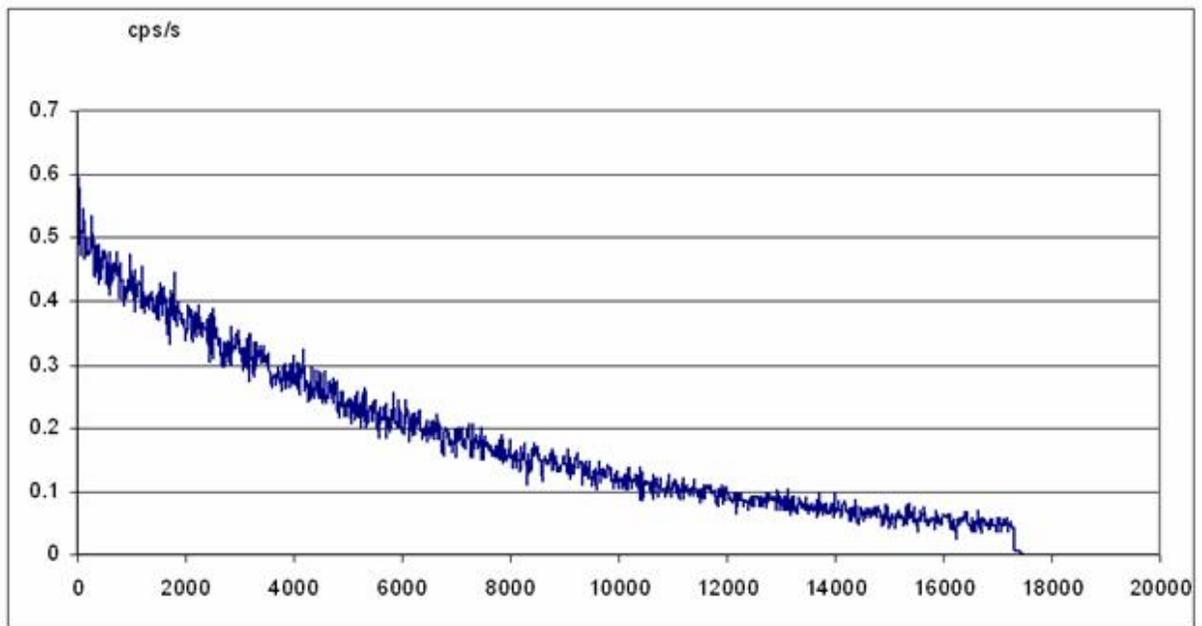

Fig. 17. Long-term measurement performed with the ionization chamber: the counting rate decrease with a good accuracy corresponds to the decay of the $^{222}$Rn ($T_0$=5500min)

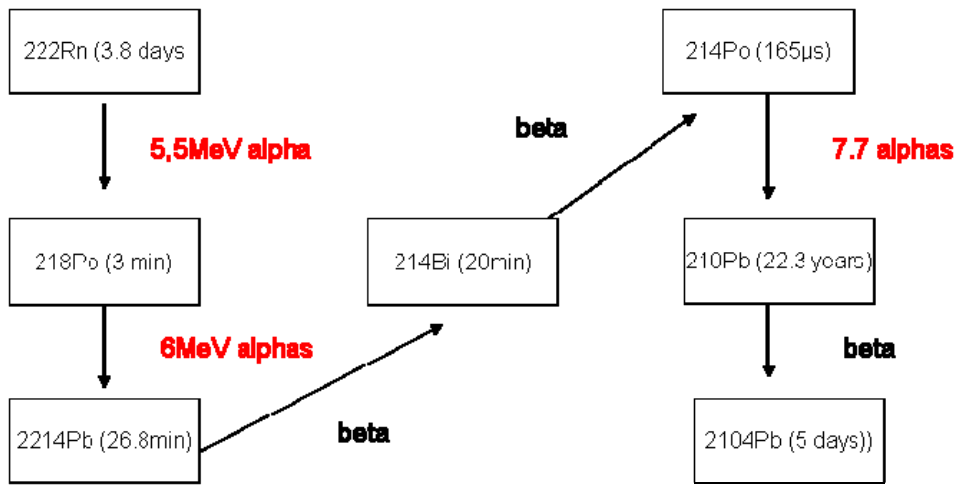

Fig.18. Chain of $^{222}$Rn decays. Our single-wire counters were mostly sensitive to alpha particles emissions

|  | Single wire proportional counters | Single wire ionization chamber | MWPC |
|---|---|---|---|
| Noise Bq/m3 | 76 | 24 | 1.2(air from a cylinder) 2 (ambient air) |
| Efficiency | 1 | 0.15 | 1 |

Table 1. Noise and efficiency for various types of wire detectors operated in air

| Time of counting (min) | Single wire proportional counters MDA | Single wire ionization chamber MDA | MWPC MDA | Atmos MDA |
|---|---|---|---|---|
| 0.2 | 1300 | 6830 | 625 |  |
| 1 | 420 | 1960 | 140 | 150 |
| 2 | 270 | 1200 | 75 |  |
| 4 | 175 | 760 | 43 |  |

Table 2. Minimum detectable activity, in Bq/m3, of Rn alpha particles for various time intervals of measurements Δt

Note that this high sensitivity was achieved after the rejection of signals having a duration <0.5ms, performed by the Lab view program. Without applying this rejection the sensitivity was 5 times lower, being only slightly worse than for commercial detectors.
The MDA of a single-wire counter for a long integration time cannot be estimated correctly since the injected Rn diffuses out from the active detector volume, the detector not being gas tight.

**IV.2 Test of the MWPC**

The main limitation in the MDA for a single-wire counter comes from the noise pulses, although their counting rate was already quite small, i.e. 0.03 per sec. The noise rate can be drastically reduced in the multi-wire design exploiting signal coincidence. This is possible because, due the short distance between the anode and the cathode wires, the number of survived electrons, which escape the attachment, $n/n_o$, is quite high: 20-30% (see paragraph 2) and signals produced by these electrons are fast (<1μs) even in ionization mode[*], which allows to use short time coincidence gate, shorter than 5 μs
Figure19 shows the gain vs. voltage as measured for inner and peripheral wires in two cases: alpha tracks oriented perpendicular and parallel to the anode wires. In both cases the alpha signal was detectable even in ionization mode (V<1.8kV).
At V>2kV multiplication starts and, due to the different geometry, the gain for the inner anode wires and for the peripheral ones (close to the cathode cylinder), was slightly different. However this difference was not relevant for the measurements.

---

[*] The MWPC design was already use in our previous work[13], but in pulse ionization mode only

At gas gain below 10, the signal shapes in Ar and air were almost identical, with a fast rise time of ~1 μs. At higher gain, an increased photon feedback after-pulse signal has been observed in Ar, but not in air. In air, up to a gain of $10^3$ the signal waveform was the same as for conventional MWPC operated with quenched gases : they had fast rise time (<1 μs ) and no after-pulses. Therefore it is possible to use these signals in coincidence mode.

Figure 20 shows the efficiency vs. voltage measured in Ar and in air. In Ar all the wires were connected to a single amplifier channel, whereas in the case of air the signals were detected in coincidence mode. It can be seen, even for alpha particles oriented almost parallel to the anode wires that the efficiency is three time lower. Theoretically for particles oriented exactly parallel to the anode wire it should be close to zero. However, since the rate of such tracks is very low:

$$sL/\{(4/3)\pi L^3\} \quad (10)$$

where s is the area of an elementary triangle cell in the wire arrangement, below one percent, the efficiency of the MWPC to alpha radiation was, with a good accuracy, 100%. MDA of the MWPC for a measurement time of 1 min was ~140Bq/m$^3$ , as for the best commercially available Rn detector (see Table 2).

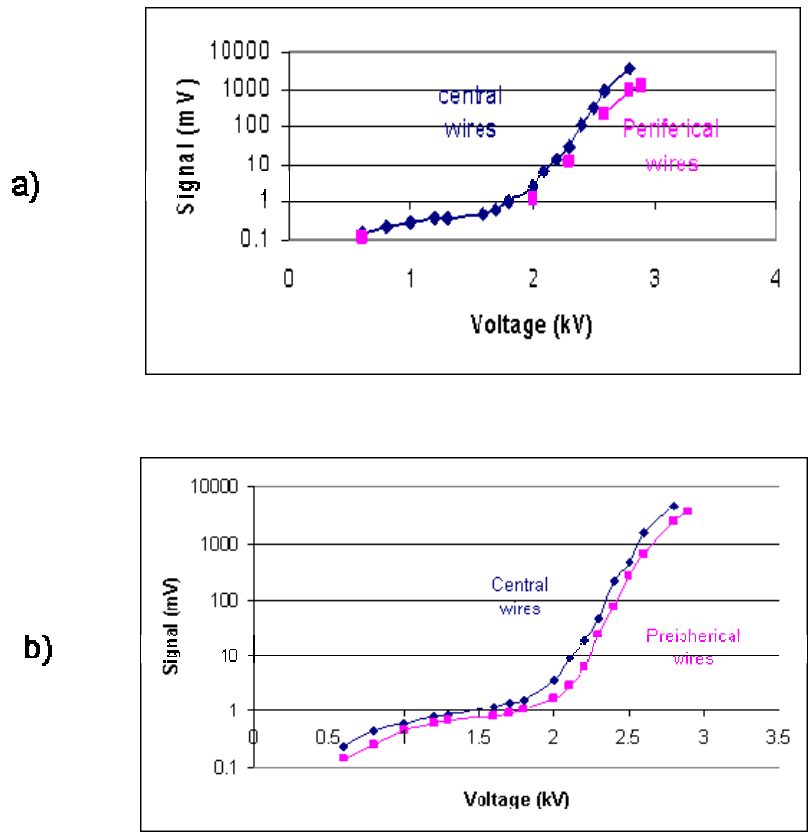

Fig 19. MWPC gain vs. voltage measured with alpha particles oriented perpendicular to anode wires(a) and parallel to them (b)

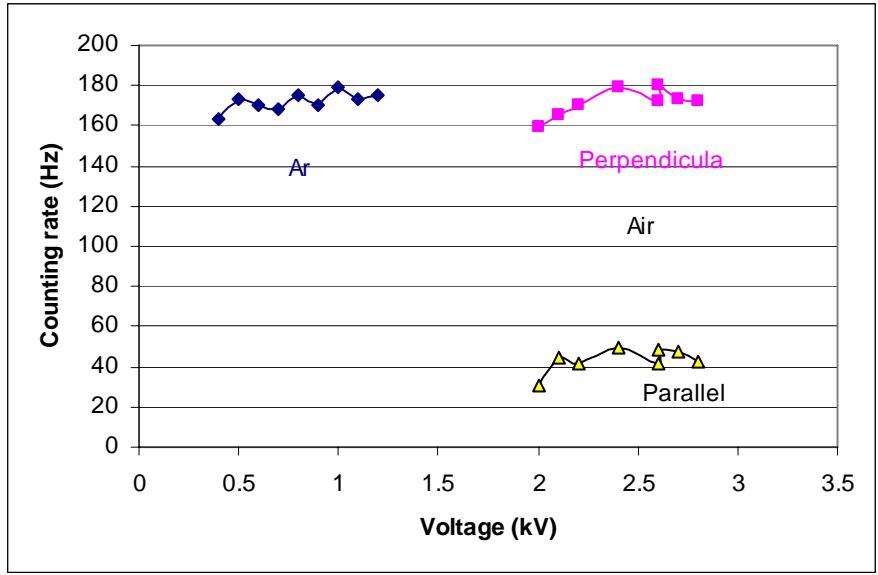

Fig. 20. Counting characteristics of the MWPC measured in Ar (all wires were connected to one amplifier) and in air for alpha tracks oriented perpendicular and parallel to the anode wire

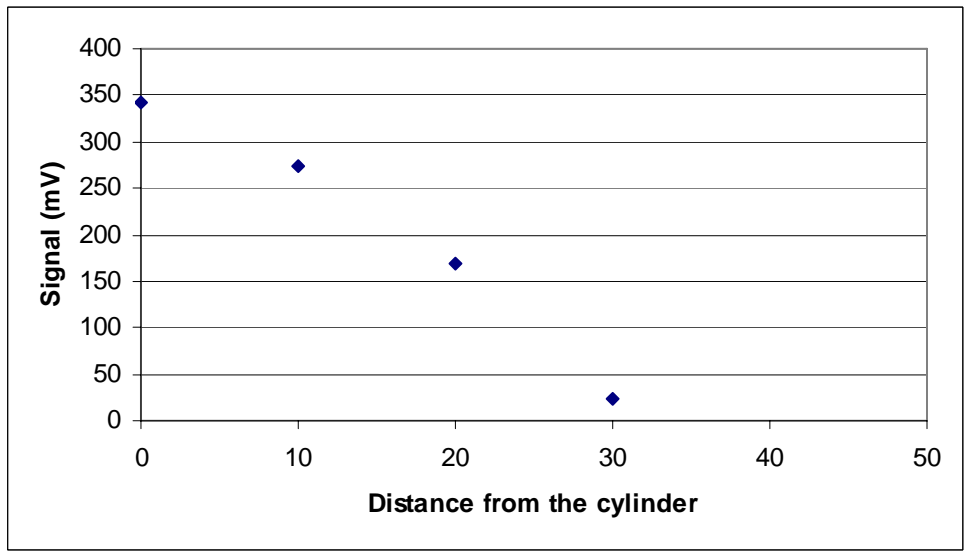

Fig. 21. Signal amplitude vs. the distance from the cathode of the MWPC measured with alpha particles oriented perpendicular to the cylinder axis. In these particular measurements all wires were connected to one amplifier

In fig. 22 is shown the efficiency for alpha particles for tracks oriented perpendicular to the anode wires, at various distances from the cathode cylinder. In these measurements, the signals were put in coincidence and therefore the efficiency becomes zero when alpha tracks entered only one section (distances > 3cm).
Finally, after exhausting tests with alpha particles, the capability of the MWPC was demonstrated under realistic conditions, detecting small concentration of Rn in air.

In fig. 23 are presented the results for one measurement, with the MWPC placed in the basement of a building. The presence of Radon was reliably detected.
More tests with the MWPC are currently in progress.

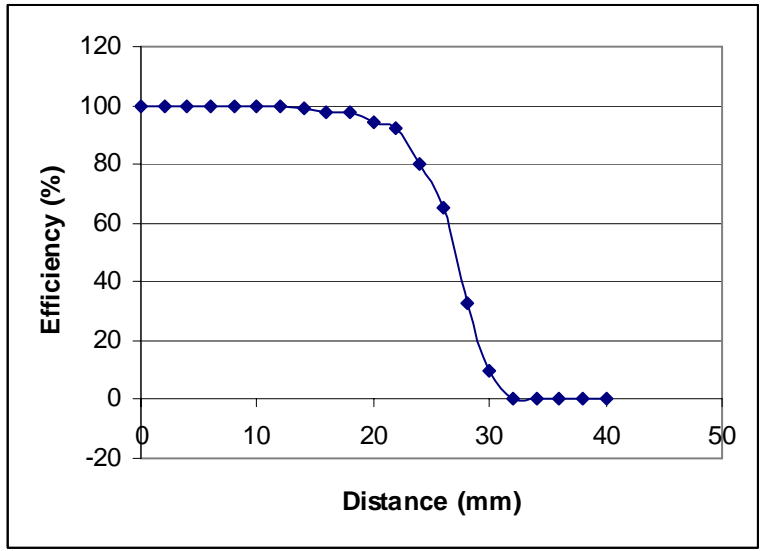

Fig. 22. Efficiency of alpha particles detection vs. the distance from the cathode cylinder when MWPC operates in signal coincidence mode

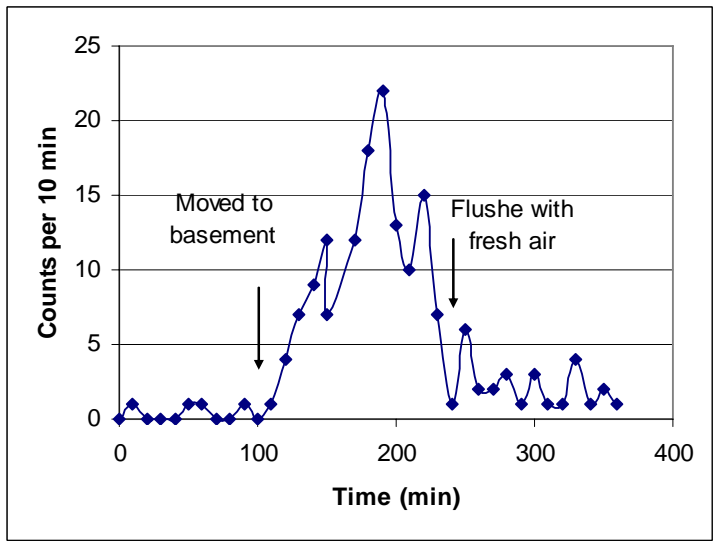

Fig.23. Performance for a small concentration of Rn in a basement. In the time interval of 0-100min, the MWPC has been operated in fresh air. At t=100min, it was moved in to the basement. At t=240min the chamber was flushed with fresh air and removed from the basement

Although this detector has a sensitivity three times higher than the single-wire sensor and equal to the best commercial one, due to its complexity we think that single-wire counters will be a more practical solution for the application to the prevision of earthquakes.

## 5. Outlook

The results reported indicate that, after optimization, the single-wire counter can operate stably at high gas gain in ambient air, even in presence of moisture, and it can be an attractive device for the online detection of Radon. The MDA of the basic design is improved by a factor ~5 by the application of the signal rejection algorithm: low and high thresholds for the signal height, and low threshold for their width, the typical minimum width being 1 ms. In a noisy environment, it was useful to perform digital filtering which eliminates signals below 100 Hz (AC 50 Hz and harmonics) and above some kHz (electrical noise). However, the use of Labview for was not essential. The advantages were the possibility to acquire raw signals, the facility of programming and making user interfaces. The mentioned above algorithms are very easy to implement in a microcontroller (microprocessor for instrumentation which costs about 3$). A prototype of such microcontroller was already built and successfully tested in the laboratory. Thus for the verification of the correlation between the sudden Rn appearance and the earthquake it is possible to use the simple, low cost and efficient detectors developed and presented in this work.

Their sensitivity in counting mode is comparable or slightly worse to the best commercial devices [29]. However, by using n detectors it could be improved by a factor $\sqrt{n}$

The assessment of multi-wire counters for this particular application is still to be completed. Although this detector has three times higher sensitivity, its design is more complicated and therefore more expensive.

Single-wire counters can also be used as cheap security devices for the monitoring of large quantities of materials and large surfaces contaminated by Po in sensitive public areas, such as airports, railway stations and so on. Due to the short path lenght of alpha tracks in air, Po can be detected only in the vicinity of the contaminated surfaces, <4cm. Due to this limitation, reliable monitoring of Po contaminations is possible only using devices with large sensitive area. Since commercial detectors are too expensive for such application, low cost single-wire counters can provide a very attractive alternative.

The schematic drawing of a prototype of a possible Po detector is presented in fig.24.

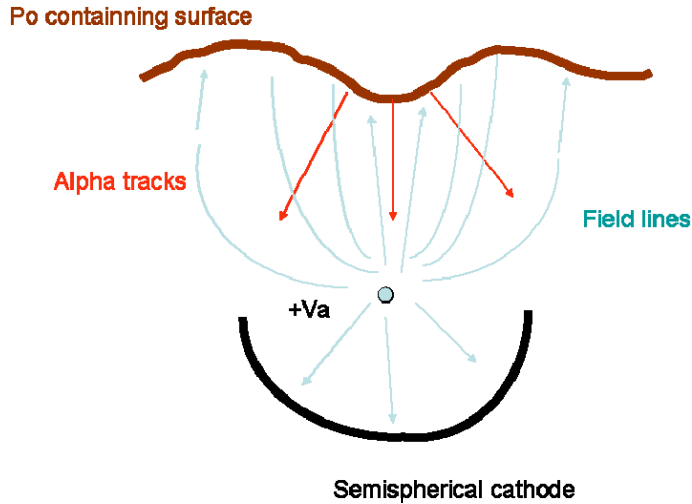

Fig.24. A schematic drawing of a single-wire counter for the monitoring of the presence of Po.

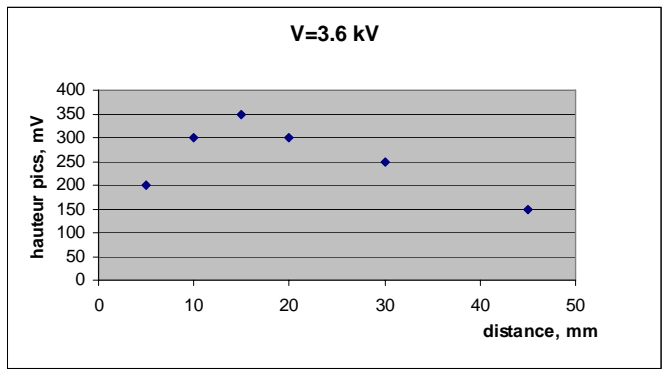

Fig. 25. Alpha particles signal amplitude as a function of the distance from the single-wire counter

It is a single-wire detector with a semi-spherical cathode cylinder. In this geometry some fraction of the field lines will end up on the investigated (grounded) surface allowing for negative ions to drift inside the detector volume where, in the high electric field near the anode wire, they will experience detachment. These free electrons liberated from the electronegative ions will then triggers Townsend avalanches. Preliminary measurements show that this single-wire counter can reliably detect alpha particles on a distance up to 40mm from the emitting surface (see fig.25). Such a simple detector does not have sufficient energy resolution and it should operate in threshold mode only. However, an array of single-wire counters can be used to trigger the first level alarm in a radioactive background monitoring system . In this particular application the measurement should be performed fast, i.e. with a small integration time, making the single wire counter very competitive

After the first warning is issued, the precise spectroscopy analysis of the suspicious area can be performed by a portable, high quality commercial device

Another possibility it is to use a MWPCs operating with gas gain, which offers 3 times higher MDA. However, we think that the single wire counters it is preferable due to its simplicity and lower cost.

## Conclusions

We have developed a robust air-filled detector for alpha particles able to operate at high gain and in presence of 100% humidity level.
Our detector has a sensitivity close to the best commercially available Radon detectors although with a worse spectral response. However, if operated in an array, it can reliably detect small variations in Rn concentration enabling to study the possible correlation between the appearance of high Rn level and forthcoming earthquakes.